\documentclass{ifacconf}
\usepackage{natbib}        
\usepackage{comment}
\usepackage{mathtools}
\mathtoolsset{showonlyrefs=true}

\usepackage{amssymb,amsmath,color} 
\usepackage{graphicx} 
\usepackage{epsfig}
\usepackage{algorithm}
\usepackage[noend]{algpseudocode}

\renewcommand{\natural}{{\mathbb{N}}} 

\newcommand{\real}{{\mathbb{R}}}

\newcommand{\map}[3]{#1: #2 \rightarrow #3}

\newcommand{\until}[1]{\{1,\ldots,#1\}}

\newcommand{\EE}{\mathcal{E}} 
\newcommand{\GG}{\mathcal{G}}\newcommand{\LL}{\mathcal{L}}

 \newcommand{\subj}{\text{subj. to}}

\newcommand{\argmax}{\mathop{\rm argmax}}

\newcommand{\nbrs}{\mathcal{N}}

\newcommand{\1}{\mathbf{1}}
\newcommand{\0}{\mathbf{0}}

\renewcommand{\inf}{\operatornamewithlimits{inf\vphantom{p}}}

\renewcommand{\lim}{\operatornamewithlimits{lim\vphantom{p}}}

\algdef{SE}[DOWHILE]{Do}{doWhile}{\hskip\algorithmicindent\algorithmicdo}[1]{\hskip\algorithmicindent\algorithmicwhile\ #1}%

\makeatletter
\newcommand{\StatexIndent}[1][3]{%
  \setlength\@tempdima{\algorithmicindent}%
  \Statex\hskip\dimexpr#1\@tempdima\relax}
\makeatother

\algnewcommand{\algorithmicgoto}{\textbf{go to }}%
\algnewcommand{\Goto}[1]{\algorithmicgoto Line~\ref{#1}}%
\algnewcommand{\Label}{\State\unskip}

\renewcommand{\algorithmicwhile}{\hskip\algorithmicindent \textbf{While:}}

\usepackage{soul} 

\newtheorem{theorem}{Theorem}[section]
\newtheorem{proposition}[theorem]{Proposition}

\newtheorem{definition}[theorem]{Definition} \newtheorem{lemma}[theorem]{Lemma}
\newtheorem{remark}[theorem]{Remark}

\newtheorem{assumption}[theorem]{Assumption}

\newcommand\oprocendsymbol{\hbox{$\square$}}
\newcommand\oprocend{\relax\ifmmode\else\unskip\hfill\fi\oprocendsymbol}
\def\eqoprocend{\tag*{$\square$}}

\usepackage{tikz}
\usetikzlibrary{shapes,arrows}

\graphicspath{{figs/}}

\def \algfullname/{Relaxation and Successive Distributed Decomposition method}
\def \algname/{RSDD}

\newcommand{\sx}[1]{\mathbf{x}_{#1}}
\newcommand{\barsx}[1]{\bar{\mathbf{x}}_{#1}}
\newcommand{\srho}[1]{\boldsymbol{\rho}_{#1}}
\newcommand{\smu}[1]{\boldsymbol{\mu}_{#1}}

\newcommand{\slambda}[1]{\boldsymbol{\lambda}_{#1}}
\newcommand{\tildeslambda}[1]{\boldsymbol{\tilde{\lambda}}_{#1}}
\newcommand{\sLambda}{\boldsymbol{\Lambda}}
\newcommand{\tildesLambda}{\boldsymbol{\tilde{\Lambda}}}

\newcommand{\IN}[1]{\ignorespaces}

\begin{document}

\begin{frontmatter}

\title{
  A Duality-Based Approach for Distributed Optimization with Coupling Constraints
  } 

\thanks{This result is part of a project that has received funding from the European Research Council (ERC)
    under the European Union's Horizon 2020 research and innovation programme
    (grant agreement No 638992 - OPT4SMART). }

\author{Ivano Notarnicola and Giuseppe Notarstefano} 

\address{Department of Engineering, Universit\`a del Salento, Lecce, Italy \\
  \texttt{$\{$ivano.notarnicola,giuseppe.notarstefano$\}$@unisalento.it.}
}

\begin{abstract}                
  In this paper we consider a distributed optimization scenario in which a set
  of agents has to solve a convex optimization problem with separable cost
  function, local constraint sets and a coupling inequality constraint. We
  propose a novel distributed algorithm based on a relaxation of the primal
  problem and an elegant exploration of duality theory. Despite its complex
  derivation based on several duality steps, the distributed algorithm has a
  very simple and intuitive structure. That is, each node solves a local version
  of the original problem relaxation, and updates suitable dual variables. We
  prove the algorithm correctness and show its effectiveness via numerical
  computations.\\[-0.6cm]
\end{abstract}


\end{frontmatter}

\section{Introduction}
\label{sec:intro}
A common set-up in large-scale optimization consists in minimizing the sum of
local cost functions, each one depending on a local variable, subject to a
constraint coupling the local decision variables. This optimization structure
arises in several concrete problems as, e.g., in resource allocation problems
(e.g., in Communications or Robotics) or energy flow optimization in smart
grids.
Solutions in a parallel, master-subproblem architecture have been known for a
while, see, e.g., \citep{bertsekas1989parallel}.
More recently \cite{tran2016fast} propose a parallel inexact dual decomposition
scheme combined with smoothing techniques for solving these separable convex
optimization problems.

In the last years a new distributed computation paradigm has been investigated
to solve optimization problems arising in a network context. Since the above
mentioned class of optimization problems has important applications in network
scenarios, proposing distributed algorithms is subject of great interest.
This class of problems has been addressed in a distributed set-up in
\citep{burger2014polyhedral}, where a cutting-plane consensus scheme is proposed
to solve the dual problem. The idea is to iteratively approximate a local
problem with linear constraints (cutting planes) and exchange the active ones
with neighboring nodes. A similar approach was applied in
\citep{burger2013non} to design a distributed model predictive control
scheme.
\cite{simonetto2016primal} propose a consensus-based distributed algorithm to
generate approximate dual solutions for this class of problems.
This distributed optimization set-up is also addressed
by~\cite{falsone2016dual}. A consensus-based proximal minimization on the dual
problem is proposed to generate a dual solution.
In these last two papers a primal recovery mechanism is proposed to obtain a
primal optimal solution.
A special coupling is considered in~\citep{notarnicola2016duality}, where a
preliminary version of the idea proposed in this paper is applied to a min-max
optimization problem for demand side management.
\cite{chang2016proximal} considers problems with a linear coupling constraint
and proposes a proximal dual consensus ADMM to solve it in a distributed way.
\cite{chang2014distributed} propose a consensus-based primal-dual perturbation
algorithm to solve optimization problems with a slightly more general
cost function (a coupling term known to all agents is allowed).
\cite{mateos2015distributed} address a class of min-max optimization problems
which are strictly related to the same problem set-up investigated in this
paper. They solve the min-max problem through a Laplacian-based saddle-point
subgradient scheme.
%

The main paper contribution is the design of a novel distributed method, based
on relaxation and duality, to solve convex optimization problems with separable
cost function and coupling constraint. The proposed algorithm is based on two
main methodological approaches. First, we consider a relaxation of the primal
problem by constraining its dual with an additional box constraint. Such a dual
is shown to have the same dual (and then primal) cost if the bound is
sufficiently large. We show that without such a relaxation the algorithmic idea
is not guaranteed to be implementable. Second, we apply duality on a series of
equivalent problems. Specifically, we generate an equivalent version of the
box-constrained dual problem in order to enforce the graph sparsity. 
By applying dual decomposition another dual problem is introduced. This final
problem has a sparse structure, so that a subgradient algorithm applied to it
turns out to be a distributed algorithm. In order to explicitly compute a local
subgradient at each node, a further duality step is performed on the local
subproblem, thus obtaining an optimization problem in the original primal
variables. Despite this lengthy and complex duality tour, the resulting
distributed algorithm has a simple and clean structure: each node finds a
primal-dual optimal solution pair of a relaxed, local version of the original
primal problem, and linearly updates some additional local dual variables.

The outline of the paper is as follows. In Section~\ref{sec:setup_preliminaries}
we formalize the distributed optimization set-up and give some preliminaries. In
Section~\ref{sec:towards} we show a first attempt to design a distributed
algorithm, which turns out not to be implementable.  In
Section~\ref{sec:algorithm} we introduce the relaxation approach, derive our
distributed optimization algorithm, and analyze it. In
Section~\ref{sec:simulations} we corroborate the theoretical results with a
numerical example.

\section{Distributed Optimization Set-up and Preliminaries}
\label{sec:setup_preliminaries}
In this section we set-up the distributed optimization framework and recall
useful preliminaries on duality.


\subsection{Distributed optimization set-up}
\label{subsec:distributed_setup}

Consider the following optimization problem
\begin{align}
\begin{split}
  \min_{\sx{1},\ldots,\sx{N}} \: & \: \textstyle\sum\limits_{i=1}^N f_i ( \sx{i} )
  \\
  \subj \: & \: \sx{i} \in X_i, \hspace{1.5cm} i \in\until{N}
  \\
  & \: \textstyle\sum\limits_{i=1}^N g_i (\sx{i}) \preceq 0
\end{split}
\label{eq:primal_problem}
\end{align}
where for all $i\in\until{N}$, the set $X_i\subseteq \real^{n_i}$ with $n_i\in\natural$,
the functions $\map{f_i}{\real^{n_i}}{\real}$ and
$\map{g_i}{\real^{n_i}}{\real^S}$ with $S\in\natural$.

\begin{assumption}
  For all $i\in\until{N}$, each function $f_i$ is convex, 
  and each $X_i$ is a non-empty, compact, convex set. Moreover, for all $s\in\until{S}$
  each component $\map{ g_{is} }{\real^{n_i}}{\real}$ of $g_i$ is a convex function.
\label{ass:primal_regualirity}
\end{assumption}

The following assumption is the well-known Slater's constraint qualification.
\begin{assumption}
  There exist $\barsx{1}\in X_1, \ldots, \barsx{N}\in X_N$ such that
  $\sum_{i=1}^N g_i (\barsx{i}) \prec \0$.
  \oprocend
\label{ass:constraints_qualification}
\end{assumption}

We consider a network of $N$ processors communicating according to a
\emph{connected, undirected} graph $\GG = (\until{N}, \EE)$, where
$\EE\subseteq \until{N} \times \until{N}$ is the set of edges. Edge $(i,j)$
models the fact that node $i$ sends information to $j$. Note that, since we
assume the graph to be undirected, for each $(i,j)\in\EE$, then also
$(j,i)\in\EE$. We denote by $|\EE|$ the cardinality of $\EE$ and by $\nbrs_i$
the set of \emph{neighbors} of node $i$ in $\GG$, i.e.,
$\nbrs_i := \left\{j \in \until{N} \mid (i,j) \in \EE \right\}$.

\subsection{Preliminaries on Optimization and Duality}
\label{subsec:optimization_duality}
Consider a constrained optimization problem, addressed as primal problem,
having the form
\begin{align}
\begin{split}
  \min_{ z \in Z } \:& \: f(z)
  \\
  \subj \: & \: g(z) \preceq 0
\end{split}
\label{eq:appendix_primal}
\end{align}
where $Z \subseteq \real^N$ is a convex and compact set,
$\map{f}{\real^N}{\real}$ is a convex function and $\map{g}{\real^N}{\real^S}$
is such that each component $\map{g_s}{\real^N}{\real}$,
$s \in \until{S}$, is a convex function.

The following optimization problem
\begin{align}
\begin{split}
  \max_{\mu} \:& \: q(\mu)
  \\
  \subj \: & \: \mu \succeq 0
\end{split}
\label{eq:appendix_dual}
\end{align}
is called the dual of problem~\eqref{eq:appendix_primal}, where
$\map{q}{\real^S}{\real}$ is obtained by minimizing with respect to $z \in Z$
the Lagrangian function $\LL (z,\mu) := f(z) + \mu^\top g(z)$, i.e.,
$q(\mu) = \min_{z \in Z} \LL(z,\mu)$. Problem~\eqref{eq:appendix_dual} is
well posed since the domain of $q$ is convex and $q$ is concave on its domain.

It can be shown that the following inequality holds
\begin{align}
  \inf_{z \in Z } \sup_{\mu \succeq 0} \LL(z, \mu) \ge \sup_{\mu\succeq 0} \inf_{z \in Z } \LL(z,\mu),
  \label{eq:appendix_weak_duality}
\end{align}
which is called weak duality.
When in~\eqref{eq:appendix_weak_duality} the equality holds, then we say
that strong duality holds and, thus, solving the primal
problem~\eqref{eq:appendix_primal} is equivalent to solving its dual
formulation~\eqref{eq:appendix_dual}. In this case the right-hand-side
problem in~\eqref{eq:appendix_weak_duality} is referred to as
\emph{saddle-point problem} of~\eqref{eq:appendix_primal}.

\begin{definition}
  A pair $(z^\star , \mu^\star)$ is called a primal-dual optimal solution of
  problem~\eqref{eq:appendix_primal} if $z^\star\in Z$ and
  $\mu^\star\succeq 0$, and $(z^\star , \mu^\star)$ is a saddle point of the
  Lagrangian, i.e.,
  \begin{align*}
    \LL (z^\star,\mu ) \le \LL (z^\star,\mu^\star) \le \LL (z,\mu^\star)
  \end{align*}
  for all $z\in Z$ and $\mu\succeq 0$.\oprocend
\label{def:primal_dual_pair}
\end{definition}

A more general min-max property can be stated. Let $Z \subseteq \real^N$ and
$W \subseteq \real^S$ be nonempty convex sets.  Let
$\phi : Z \times W \to \real$, then the following inequality
\begin{align*}
  \inf_{z\in Z} \sup_{w \in W}  \phi (z,w) \ge \sup_{w \in W} \inf_{z\in Z} \phi (z,w)
\end{align*}
holds true and is called the \emph{max-min} inequality. When the equality holds, then
we say that $\phi$, $Z$ and $W$ satisfy the \emph{strong max-min} property or the
\emph{saddle-point} property.

The following theorem gives a sufficient condition for the strong max-min property
to hold.
\begin{proposition}[{\citep[Propositions~4.3]{bertsekas2009min}}]\mbox{}\\[-1.6em]

  Let $\phi$ be such that (i) $\map{\phi (\cdot,w)}{Z}{\real}$ is convex and closed
  for each $w \in W$, and (ii) $\map{-\phi(z,\cdot)}{W}{\real}$ is convex and closed for
  each $z \in Z$.
  Assume further that $W$ and $Z$ are convex compact sets. 
  Then
    $\sup_{w \in W} \inf_{z \in Z} \phi (z,w) = \inf_{z \in Z} \sup_{w \in W}  \phi (z,w)$
  and the set of saddle points is nonempty and compact.~\oprocend
  \label{prop:saddle_point}
\end{proposition}



\section{Towards A Distributed Optimization Algorithm}
\label{sec:towards}
In this section we provide a first attempt to design a duality-based distributed
algorithm to solve problem~\eqref{eq:primal_problem}, but then show that in general it is not guaranteed 
to be implementable.


\subsection{A First Dual Problem Derivation}
\label{subsec:dual_problem_derivation}

We start by deriving the equivalent dual problem of~\eqref{eq:primal_problem} as
formally stated in the next lemma.
\begin{lemma}
  Let Assumptions~\ref{ass:primal_regualirity} and~\ref{ass:constraints_qualification} hold.
  The optimization problem
  \begin{align}
  \begin{split}
    \max_{\smu{} \in \real^S} \: & \: \textstyle\sum\limits_{i=1}^N q_i (\smu{})
    \\
    \subj \: & \: \smu{} \succeq \0
  \end{split}
  \label{eq:dual_problem}
  \end{align}
  with
  \begin{align}
    q_i (\smu{}) := \min_{ \sx{i} \in X_i} \Big( f_i ( \sx{i} ) + \smu{}^\top g_i( \sx{i} ) \Big),
    \label{eq:qi_definition}
  \end{align}
  for all $i\in\until{N}$, is the dual of problem~\eqref{eq:primal_problem}.
  Moreover, both problems~\eqref{eq:primal_problem} and~\eqref{eq:dual_problem} have finite optimal 
  cost, respectively $f^\star$ and $q^\star$, and strong duality holds, i.e.,
    $f^\star = q^\star$.~\oprocend
  %
  \label{lem:strong_duality}
\end{lemma}
The proof of this lemma is omitted for the sake of space, but can be shown by
using classical methods from duality theory.

\begin{remark}
  Since $q(\smu{}) = \sum_{i=1}^N q_i (\smu{})$ is the dual function
  of~\eqref{eq:primal_problem}, it is concave on its convex domain, which can be
  shown to be the entire $\smu{} \succeq \0$.\oprocend
\label{rem:dual_properties}
\end{remark}


\subsection{Tentative Distributed Dual Subgradient}
\label{subsec:tentative_distributed}

We focus now on the solution of problem~\eqref{eq:dual_problem}.  In order to
make problem~\eqref{eq:dual_problem} amenable for a distributed solution, we
need to enforce a sparsity structure. To this end, we introduce copies of the
common optimization variable $\smu{}$, and enforce coherence constraints having
the sparsity of the connected graph $\GG$, thus obtaining
\begin{align}
\begin{split}
  \max_{\smu{1}, \ldots, \smu{N} } \: & \: \textstyle\sum\limits_{i=1}^N q_i (\smu{i} )
  \\
  \subj \: & \: \smu{i} \succeq \0, \hspace{1.1cm}  i \in\until{N}
  \\
  & \: \smu{i} = \smu{j}, \hspace{0.9cm} (i,j) \in \EE.
\end{split}
\label{eq:dual_problem_copies}
\end{align}
Being problem~\eqref{eq:dual_problem_copies} an equivalent version of
problem~\eqref{eq:dual_problem}, it has the same optimal cost $q^\star$. 

On this problem we would like to use a dual decomposition approach with the aim
of obtaining a distributed algorithm. That is, the tentative idea is to derive
the dual of problem~\eqref{eq:dual_problem_copies} and apply a dual subgradient
algorithm.

We start deriving the dual of~\eqref{eq:dual_problem_copies} 
by dualizing only the coherence constraints. Thus, we write the partial Lagrangian
\begin{align}
\begin{split}
  \LL_2(\smu{1},\ldots,  \smu{N}, \sLambda ) 
 \! = \! 
 \textstyle\sum\limits_{i=1}^N \! \Big( q_i( \smu{i} ) \!+ \!\! \textstyle\sum\limits_{j\in\nbrs_i} \slambda{ij}^\top (\smu{i} - \smu{j} ) \! \Big)
\end{split}
\label{eq:lagrangian_dual_copies}
\end{align}
where $\sLambda\in \real^{S\cdot |\EE|}$ is the vector stacking each Lagrange
multiplier $\slambda{ij} \in \real^S$, with $(i,j)\in \EE$, associated to the
constraint $\smu{i} - \smu{j} = \0$.

Since the communication graph $\GG$ is undirected and connected, we can exploit
the symmetry of the constraints. In fact, for each $(i,j)\in\EE$ we also have
$(j,i) \in\EE$, and, expanding all the terms in
\eqref{eq:lagrangian_dual_copies}, for given $i$ and $j$, we always have both
the terms $\slambda{ij}^\top (\smu{i} - \smu{j} )$ and
$\slambda{ji}^\top (\smu{j} - \smu{i} )$.
%
Thus, after some simple algebraic manipulations, we get
\begin{align*}
\begin{split}
  \LL_2(\smu{1}, \ldots, \smu{N}, \sLambda ) = 
  \textstyle\sum\limits_{i=1}^N \Big( q_i( \smu{i} ) + \smu{i}^\top \!\! \sum\limits_{j\in\nbrs_i} (\slambda{ij} - \slambda{ji})  \Big),
\end{split}
\end{align*}
which is separable with respect to $\smu{i}$, $i\in\until{N}$.
Thus, the dual function of~\eqref{eq:dual_problem_copies} is
\begin{align*}
  \eta( \sLambda ) & \!=
    \!\! \sup_{\smu{1} \succeq \0, \ldots,\smu{N} \succeq \0}\!\!
    \LL_2(\smu{1}, \ldots, \smu{N},  \sLambda )
    \\
    & = \textstyle\sum\limits_{i=1}^N \eta_i( \{ \slambda{ij},\slambda{ji} \}_{j\in\nbrs_i})
\end{align*}
where, for all $i\in\until{N}$,
\begin{align}
  \eta_i \big( \{ \slambda{ij},\slambda{ji} \}_{j\in\nbrs_i} \big) \! := \!\!
    \sup_{  \smu{i} \succeq \0 }  \! 
      \Big( \! q_i (\smu{i})   \! + \! 
        \smu{i}^{\!\top} \! \!  \textstyle\sum\limits_{j\in\nbrs_i} \! (\slambda{ij}  \! -  \!  \slambda{ji} )
    \! \Big).
\label{eq:eta_definition}
\end{align}
Finally, by denoting the domain of $\eta$ as
\begin{align*}
  D_{\sLambda } = \big\{ \sLambda \in \real^{S \cdot |\EE|} \mid \eta( \sLambda ) < +\infty \big\},
\end{align*}
we can state the dual of problem~\eqref{eq:dual_problem_copies} as
\begin{align}
  \min_{ \sLambda \in D_{ \sLambda} } \eta( \sLambda ) := 
  \textstyle\sum\limits_{i=1}^N \eta_i \big ( \{\slambda{ij},\slambda{ji}\}_{j\in\nbrs_i} \big).
\label{eq:dual_dual}
\end{align}

Since problem~\eqref{eq:dual_dual} is the dual of~\eqref{eq:dual_problem_copies}
we recall, \citep[Section~8.1]{bertsekas2003convex}, how to compute the 
components of a subgradient\footnote{A vector
  $\widetilde{\nabla} f( z ) \in\real^N$ is called a subgradient of the convex
  function $f$ at $z\in\real^N$ if
  $f(y) \ge f(z) + \widetilde{\nabla} f( z ) (y - z)$ for all $y\in\real^N$.} of
$\eta$ at a given $\sLambda \in D_{\sLambda}$. That is, it holds
\begin{align}
\frac{\tilde \partial
   \eta ( \sLambda ) }{ \partial \slambda{ij} }
  = \smu{i}^\star - \smu{j}^\star,
\label{eq:eta_subgradient}
\end{align}
where $\frac{\tilde \partial \eta ( \cdot )}{\partial \slambda{ij}}$ denotes the component
associated to the variable $\slambda{ij}$ of a subgradient of $\eta$, and
\begin{align}
  \smu{k}^\star \in \argmax_{ \smu{k} \succeq \0 } \Big( q_k( \smu{k} ) +
  \smu{k}^\top \textstyle\sum\limits_{h\in\nbrs_k} (\slambda{kh} - \slambda{hk} ) \Big),
\label{eq:mustar_k}
\end{align}
for $k=i,j$.

It is worth noting that since $\sLambda \in D_{\sLambda}$, then each
$\eta_k \big( \{ \slambda{kh},\slambda{hk} \}_{h\in\nbrs_k} \big) $ defined 
in~\eqref{eq:eta_definition} is finite and therefore a $\smu{k}^\star$ 
in~\eqref{eq:mustar_k} exists.

A viable solution to solve problem~\eqref{eq:dual_problem_copies} 
is to apply a dual projected subgradient method, which can be stated as:
\begin{itemize}
\item[(S1)] for each $i\in\until{N}$, collect $\slambda{ji}(t)$, $j \in \nbrs_i$, and
  compute a subgradient $\smu{i}(t+1)$ by solving
  \begin{align}
    \max_{ \smu{i} \succeq \0 } \Big( q_i( \smu{i} ) +
    \smu{i}^\top \!\! 
    \textstyle\sum\limits_{j\in\nbrs_i} (\slambda{ij}(t) - \slambda{ji}(t)) \Big);
    \label{eq:dual_subgradient}
  \end{align}

\item[(S2)] for each $i\in\until{N}$, exchange with neighboring nodes the updated
  $\smu{j}(t+1)$, $j \in \nbrs_i$, and compute $\tildeslambda{ij}$,
  $j\in\nbrs_i$, via
  \begin{align}
    \tildeslambda{ij} (t \!+\! 1) = \slambda{ij}(t) - 
      \gamma(t) (\smu{i}(t \!+\! 1) \!-\! \smu{j}(t \!+\! 1)),
    \label{eq:lambda_descent}
  \end{align}
  with $\gamma(t)$ a suitable step-size;

\item[(S3)] update $\slambda{ij}$, with $(i,j) \in \EE$, via
  \begin{align}
    \sLambda (t+1) = \Big[ \tildesLambda (t+1) \Big]_{D_{\sLambda}}
  \end{align}
  \!\!where $[\cdot]_{D_{\sLambda}}$ denotes the Euclidean projection onto $D_{\sLambda}$.
\end{itemize}

At this point it is worth discussing algorithm (S1)-(S3) since there are two
main issues that need to be addressed if we want to turn it into an implementable
distributed algorithm.
First of all, it is interesting to notice that the cost function of
problem~\eqref{eq:dual_dual} is separable and each term $\eta_i$ depends only on
neighboring variables $\slambda{ij}$ and $\slambda{ji}$ with $j \in\nbrs_i$. As
a consequence steps (S1) and (S2) have a distributed structure.
However, it is not clear how to implement step (S1) since function $q_i$
in~\eqref{eq:dual_subgradient} is not given explicitly.
Second, one should characterize $D_{\sLambda}$ and its projection
$[\,\cdot\,]_{D_{\sLambda}}$.

In the next section we will address these two issues and propose a distributed
optimization algorithm that solves
problem~\eqref{eq:primal_problem}. Specifically, we will further explore the
dual subgradient algorithm in order to have an implementable version of step
(S1). Moreover, we will propose a strategy to extend the domain $D_{\sLambda}$
to be the entire $\real^{S \cdot |\EE|}$ and thus remove the projection step
(S3).


\section{\algfullname/}
\label{sec:algorithm}
In this section we propose a strategy to overcome the issues raised up in the
previous one. We introduce a relaxation approach, then present our
distributed algorithm and finally state its convergence in objective value.
%


\subsection{Relaxation Approach}
\label{subsec:relaxation}

We start by introducing the optimization problem
\begin{align}
  \begin{split}
    \max_{\smu{} \in \real^S} \: & \: \textstyle\sum\limits_{i=1}^N q_i (\smu{})
    \\
    \subj \: & \: \0 \preceq \smu{} \preceq M \1
  \end{split}
  \label{eq:dual_problem_relaxed}
\end{align}
with $M>0$ and $\1 = [1,\ldots,1]^\top$. This problem is very similar
to~\eqref{eq:dual_problem}, but an additional constraint, namely
$\smu{} \preceq M \1$, has been added to make the constraint set compact.

It is worth noting that, in light of Lemma~\ref{lem:strong_duality}, the optimal
cost of~\eqref{eq:dual_problem} is a finite value $q^\star$ and, thus, is attained at
some $\smu{}^\star \in \real^S$, such that $q( \smu{}^\star ) = q^\star$.
The next result establishes the relation between
problem~\eqref{eq:dual_problem_relaxed} and problem~\eqref{eq:dual_problem}.
\begin{lemma}
  Let $\smu{}^\star$ be an optimal solution of problem~\eqref{eq:dual_problem} and
  $M>0$ be such that $M > \| \smu{}^\star \|_\infty$.
  Then, problem~\eqref{eq:dual_problem_relaxed} and problem~\eqref{eq:dual_problem}
  have the same optimal cost, namely $q^\star = f^\star$.
  Moreover, $\smu{}^\star$ is an optimal solution also of~\eqref{eq:dual_problem_relaxed}.\oprocend
\label{lem:relaxation_equivalence}
\end{lemma}

Next we show that problem~\eqref{eq:dual_problem_relaxed} is the dual of a
\emph{relaxed} version of problem~\eqref{eq:primal_problem}. In fact, let us
consider the following optimization problem
\begin{align}
\begin{split}
  \min_{ \sx{1},\ldots,\sx{N}, \srho{} } \: & \: \textstyle\sum\limits_{i=1}^N f_i ( \sx{i} ) + M \1^\top \srho{} 
  \\
  \subj \: & \: \srho{} \succeq \0, \:\: \sx{i} \in X_i, \: i \in\until{N}
  \\
  & \: \textstyle\sum\limits_{i=1}^N g_i (\sx{i}) \preceq \srho{}.
\end{split}
\label{eq:primal_problem_relaxed}
\end{align}
This problem is a relaxation of problem~\eqref{eq:primal_problem} in which we
allow for the violation of the coupling constraint, but at the same time we also
penalize the magnitude of the violation.

To show that problem~\eqref{eq:dual_problem_relaxed} is the dual of 
problem~\eqref{eq:primal_problem_relaxed}, consider the (partial) Lagrangian of~\eqref{eq:primal_problem_relaxed}
\begin{align*}
  \LL_3 ( \sx{1},\ldots,\sx{N}, \srho{}, \smu{} ) 
  \! & = \!\! \textstyle\sum\limits_{i=1}^N \! f_i (\sx{i}) \!\!+\!\! M \1^{\!\top } \!\! \srho{} \!+\! \smu{}^{\!\top } \! 
  \Big( \! \textstyle\sum\limits_{i=1}^N \! g_i (\sx{i})\! -\!\! \srho{} \! \Big)
  \\
  & \!\! = \!\! \textstyle\sum\limits_{i=1}^N \!\Big( \! f_i (\sx{i}) \!+\! \smu{}^{\!\top } g_i (\sx{i}) \!\Big) \!+\! {\srho{}}^{\!\top } \!(M \1 \! - \! \smu{}).
\end{align*}
Then the dual function is
\begin{align*}
  \! \! q_R(\smu{}) \! &=
    \min_{ \sx{1}\in X_1,\ldots,\sx{N}\in X_N, \srho{}\succeq \0 } \LL ( \sx{1},\ldots,\sx{N}, \srho{}, \smu{} )
  \\ & =
  \begin{cases}
    \textstyle\sum\limits_{i=1}^N \displaystyle \min_{ \sx{i}\in X_i } \Big( f_i (\sx{i}) + \smu{}^{\!\top } g_i
    (\sx{i}) \Big) & \text{if} \; M \1 - \smu{} \succeq \0
    \\
    -\infty & \text{otherwise}
  \end{cases}
  \\ & =
  \begin{cases}
    q(\smu{}) & \text{if} \; M \1 - \smu{} \succeq \0
    \\
    -\infty & \text{otherwise}.
  \end{cases}
\end{align*}
The maximization of the dual function $q_R(\smu{})$ over $\smu{}\succeq \0$
turns out to be the maximization of $q(\smu{})$ over
$\0 \preceq \smu{} \preceq M\1$, which is
problem~\eqref{eq:dual_problem_relaxed}.

At this point, we try to solve problem~\eqref{eq:dual_problem_relaxed} instead of the ``original''
dual problem~\eqref{eq:dual_problem} by using the procedure described in Section~\ref{sec:towards}.

In order to make problem~\eqref{eq:dual_problem_relaxed} amenable for a
distributed computation, we can rewrite it into an equivalent form. To this end,
we introduce copies of the common optimization variable $\smu{}$ and coherence
constraints having the sparsity of the connected graph $\GG$, thus obtaining
\begin{align}
\begin{split}
  \max_{\smu{1}, \ldots, \smu{N} } \: & \: \textstyle\sum\limits_{i=1}^N q_i (\smu{i} )
  \\
  \subj \: & \: \0 \preceq \smu{i} \preceq M\1, \hspace{0.83cm} i \in\until{N}
  \\
  & \: \smu{i} = \smu{j}, \hspace{1.65cm} (i,j) \in \EE.
\end{split}
\label{eq:dual_problem_relaxed_copies}
\end{align}
This problem has the same structure of problem~\eqref{eq:dual_problem_copies}
with additional constraints $\smu{i} \preceq M\1$, $i \in\until{N}$.

We retrace the same derivation already developed for the ``non-relaxed'' case
discussed in the previous section. Thus, we apply a dual decomposition approach
with the aim of obtaining a distributed algorithm. Differently from the previous
section, we will see that there is no restriction on the domain of $\eta$ and
thus a dual subgradient algorithm applied to
problem~\eqref{eq:dual_problem_relaxed_copies} turns out to be a distributed
optimization algorithm.

Since many objects in the derivation are the counterpart of the ones in the
previous section, we slight abuse notation and use the same symbols for formally
different objects.

First, we notice that when the box constraints
$\0\preceq\smu{i}\preceq M\1$, $i\in\until{N}$, are not dualized, the partial Lagrangian
of~\eqref{eq:dual_problem_relaxed_copies} 
is~\eqref{eq:lagrangian_dual_copies}, thus the dual function of 
problem~\eqref{eq:dual_problem_relaxed_copies} is 
\begin{align}
\eta( \sLambda ) = \textstyle\sum\limits_{i=1}^N \eta_i( \{ \slambda{ij},\slambda{ji}
\}_{j\in\nbrs_i})
\label{eq:eta_relaxed}
\end{align}
with
\begin{align}
  \eta_i \big( \{ \slambda{ij},\slambda{ji} \}_{j\in\nbrs_i} \big) \! := \!
    \sup_{  \0 \preceq \smu{i} \preceq M\1 }  \! 
      \Big( \! q_i (\smu{i})   \! + \! 
        \smu{i}^\top \! \!  \textstyle\sum\limits_{j\in\nbrs_i}(\slambda{ij}  \! -  \!  \slambda{ji} )
    \! \Big)
\label{eq:eta_i_relaxed_definition}
\end{align}
for all $i\in\until{N}$.
Notice that $\eta_i$ introduced in~\eqref{eq:eta_i_relaxed_definition} is different
from the one introduced in~\eqref{eq:eta_definition} due to the presence of 
the constraint $\smu{i} \preceq M\1$.
Finally, by denoting the domain of $\eta$ as
  $D_{\sLambda } = \big\{ \sLambda \in \real^{S \cdot |\EE|} \mid \eta( \sLambda ) < +\infty \big\}$,
%
the dual of problem~\eqref{eq:dual_problem_relaxed_copies} is
\begin{align}
  \min_{ \sLambda \in D_{\sLambda } } \eta ( \sLambda  ) := 
  \textstyle\sum\limits_{i=1}^N
  \eta_i \big( \{\slambda{ij},\slambda{ji}\}_{j\in\nbrs_i} \big),
\label{eq:dual_relaxed_dual}
\end{align}

In the next lemma we characterize the domain for
problem~\eqref{eq:dual_relaxed_dual}.

\begin{lemma}
  The domain $D_{\sLambda}$ of $\eta$ in~\eqref{eq:eta_relaxed} is
  $\real^{S \cdot |\EE|}$, thus optimization
  problem~\eqref{eq:dual_relaxed_dual} is unconstrained.~\oprocend
\label{lem:dual_relaxed_domain}
\end{lemma}


\subsection{Distributed Algorithm Derivation}
\label{subsec:distributed_alg_derivation}

Next, we introduce the distributed duality-based algorithm.
As already shown in the previous section, the separability structure of the dual
function $\eta$ gives rise to a sparse computation. Specifically, the
subgradients of $\eta$ can be computed as in~\eqref{eq:eta_subgradient}, where
$\smu{i}^\star$ and $\smu{j}^\star$ are not the ones
in~\eqref{eq:mustar_k}, but are given by
\begin{align*}
  \smu{k}^\star \in \argmax_{ \0 \preceq \smu{k} \preceq M\1 } 
  \Big( q_k( \smu{k} ) +
  \smu{k}^\top \textstyle\sum\limits_{h\in\nbrs_k} (\slambda{kh} - \slambda{hk}) \Big),
\end{align*}
for $k=i,j$.

The dual subgradient algorithm for
problem~\eqref{eq:dual_problem_relaxed_copies} can be summarized as follows. For
each node $i\in\until{N}$:
\begin{itemize}
\item[(R1)]\label{item:s1} receive $\slambda{ji}(t)$, $j \in \nbrs_i$, and
  compute a subgradient $\smu{i}(t+1)$ by solving
  \begin{align}
    \max_{ \0 \preceq \smu{i} \preceq M\1 } \Big ( q_i( \smu{i} ) +
    \smu{i}^\top \!\! \textstyle\sum\limits_{j\in\nbrs_i} (\slambda{ij}(t) - \slambda{ji}(t)) \Big )
  \label{eq:dual_relaxed_subgradient}
  \end{align}
\item[(R2)]\label{item:s2} exchange with neighbors the updated $\smu{j}(t+1)$, $j \in
  \nbrs_i$, and update $\slambda{ij}$, $j\in\nbrs_i$, via
\begin{align*}
  \slambda{ij} (t \!+\! 1) = \slambda{ij}(t) - \gamma(t) (\smu{i}(t \!+\! 1) \!-\! \smu{j}(t \!+\! 1)),
\end{align*}
with $\gamma(t)$ a suitable step-size.
\end{itemize}

Notice that (R1)-(R2) is a distributed algorithm. Here in fact, in light of
Lemma~\ref{lem:dual_relaxed_domain}, no projection is needed (differently from
the previous steps (S1)-(S3)).
Indeed, in this case, as shown in Lemma~\ref{lem:dual_relaxed_domain},
problem~\eqref{eq:dual_relaxed_subgradient} has a solution for every
$\slambda{ij}$ and $\slambda{ji}$ due to the compactness of the constraint set
$\0 \preceq \smu{i} \preceq M\1$.

However, we want to stress, once again, that the algorithm is \emph{not}
implementable as it is written, since functions $q_i$ are still not available in closed form.
In the following, we propose a technique to explicit
step~\eqref{eq:dual_relaxed_subgradient} for computation and thus obtain our
distributed algorithm.

We can rephrase problem~\eqref{eq:dual_relaxed_subgradient} by
plugging in the definition of $q_i$, given in~\eqref{eq:qi_definition}, thus
obtaining the following max-min optimization problem
\begin{align}
  \max_{ \0 \preceq \smu{i} \preceq M\1 }
  \min_{\sx{i} \in X_i} \!\!
  \Big( f_i ( \sx{i} ) \!+\! \smu{i}^\top \! \big( 
    g_i( \sx{i} ) \!+ \! \!\! \textstyle\sum\limits_{j\in\nbrs_i} \! ( \slambda{ij}(t) \!-\! \slambda{ji}(t) ) 
  \big) \! \Big).
\label{eq:maxmin}
\end{align}
%


The next lemma allows us to recast problem~\eqref{eq:maxmin} in terms of the
primal-dual optimal solution pair of a suitable optimization problem.
\begin{lemma}
  Max-min optimization problem~\eqref{eq:maxmin} is the saddle point problem
  associated to the problem
  \begin{align}
    \begin{split}
    \min_{\sx{i}, \srho{i}} \: & \: f_i (\sx{i}) + M \1^\top \srho{i}
    \\
    \subj \: & \: \srho{i} \succeq \0, \:\: \sx{i} \in X_i
    \\
    & \: g_i (\sx{i} ) + \textstyle\sum\limits_{j\in\nbrs_i}  \big( \slambda{ij}(t) -
      \slambda{ji}(t) \big) \preceq \srho{i}.
    \end{split}
  \label{eq:alg_minimization_lemma}
  \end{align}
  Moreover, a finite primal-dual optimal solution pair
  of~\eqref{eq:alg_minimization_lemma}, call it
  $( ( \sx{i} (t+1), \srho{i} (t+1) ), \smu{i}(t+1) )$, does exist and
  $( \sx{i} (t+1), \smu{i}(t+1) )$ is a solution of~\eqref{eq:maxmin}.
  \oprocend
\label{lem:dual_minmax_equivalence}
\end{lemma}%


We are now ready to present our \algfullname/ (\algname/).
Informally, the algorithm consists of a two-step procedure.  First, each node
$i\in\until{N}$ stores a set of variables (($\sx{i}$, $\srho{i}$), $\smu{i}$)
obtained as a primal-dual optimal solution pair of problem
\eqref{eq:alg_minimization_lemma}, which is a local version of the relaxed
centralized problem. The coupling with the other nodes in the original (hard)
formulation is replaced by a relaxed constraint depending on neighboring
variables $\slambda{ij}$, $j\in\nbrs_i$ and on $\srho{i}$. The variables
$\slambda{ij}$, $j\in\nbrs_i$, are updated in the second step according to a
suitable linear law, which weights the difference of neighboring $\smu{i}$.
Nodes use a suitable step-size denoted by $\gamma(t)$ and can initialize the
variables $\slambda{ij}$, $j\in\nbrs_i$ to arbitrary values.
In the next table we formally state the distributed algorithm from the
perspective of node $i$.
\begin{algorithm}
\renewcommand{\thealgorithm}{}
\floatname{algorithm}{Distributed Algorithm}

  \begin{algorithmic}[0]
    \Statex \textbf{Processor states}: $\sx{i}$, $\srho{i}$, $\smu{i}$ and $\slambda{ij}$ for $j\in\nbrs_i$

    \Statex \textbf{Evolution}:

      \StatexIndent[0.5] \textbf{Gather} $ \slambda{ji}(t)$ from $j\in\nbrs_i$

      \StatexIndent[0.5] \textbf{Compute} $((\sx{i}(t+1), \srho{i}(t+1)),\smu{i}(t+1))$ as a
        primal-dual optimal solution pair of
      \begin{align}
      \begin{split}
        \min_{\sx{i},\srho{i} } \: & \: f_i ( \sx{i} ) + M  \1^\top \srho{i}
        \\
        \subj \: & \: \srho{i} \succeq \0,\: \: \sx{i} \in X_i
        \\
        & \:  g_i (\sx{i}) + \textstyle\sum\limits_{j\in\nbrs_i} \Big( \slambda{ij}(t) - \slambda{ji}(t) \Big) \preceq \srho{i}
      \end{split}
      \label{eq:alg_minimization}
      \end{align}

      \StatexIndent[0.5] \textbf{Gather} $ \mu_j (t+1)$ from $j\in\nbrs_i$

      \StatexIndent[0.5] \textbf{Update} for all $j\in\nbrs_i$
      \begin{align}
        \slambda{ij} (t+1) & = \slambda{ij} (t) - \gamma(t) \Big( \smu{i} (t+1) - \smu{j} (t+1) \Big)
      \label{eq:alg_update}
      \end{align}

  \end{algorithmic}
  \caption{\algname/}
  \label{alg:algorithm}
\end{algorithm}

We point out that here, once again, we slightly abuse notation since 
in (R1)-(R2) we use $\smu{i}(t)$ as in \algname/, but we have not
proven the equivalence yet. Since we will prove it in the next, we preferred not
to further overweight the notation.


\subsection{Convergence}
\label{subsec:analysis}

In this section we give the converge results  of the proposed distributed algorithm.
The proofs will be provided in a forthcoming document.
To this end we start by giving some results which will act as building blocks to prove
the convergence in objective value of \algname/ algorithm.

\begin{lemma}
  For any $\sLambda(t) \in \real^{S\cdot|\EE|}$ with components
  $\slambda{ij}(t) \in \real^S$, $(i,j)\in \EE$, any subgradient of the function
  $\eta$ at $\sLambda(t)$ is bounded.~\oprocend
\label{lem:bounded_subgradient}
\end{lemma}

\begin{lemma}
  Let $\smu{}^\star$ be an optimal solution of problem~\eqref{eq:dual_problem}
  and $M>0$ be such that $M > \| \smu{}^\star \|_\infty$.

  Problem~\eqref{eq:dual_relaxed_dual}, which is the dual of
  problem~\eqref{eq:dual_problem_relaxed_copies}, has a bounded optimal cost,
  call it $\eta^\star$, and strong duality holds. Moreover,
    $\eta^\star = f^\star$,
  with $f^\star$ the optimal solution of primal problem~\eqref{eq:primal_problem}.\oprocend
  \label{lem:dual_dual}
\end{lemma}
%


We are now ready to state the main result of the paper, namely the convergence
of the \algname/ distributed algorithm. First, we need the following
assumption on the step-size.
\begin{assumption}
\label{ass:step-size}
  The sequence $\{ \gamma(t)\}$, with $\gamma(t) \ge 0$ for all $t\ge 0$,
  satisfies the diminishing condition
  \begin{align}\notag
    \textstyle \lim\limits_{t\to \infty} \gamma(t) = 0, \:\:
    \sum\limits_{t=1}^{\infty} \gamma(t) = \infty, \:\:
    \sum\limits_{t=1}^{\infty} \gamma(t)^2 < \infty. \eqoprocend
  \end{align}
\end{assumption}

\begin{theorem}
  Let $\smu{}^\star$ be an optimal solution of problem~\eqref{eq:dual_problem}
  and $M>0$ be such that $M > \| \smu{}^\star \|_\infty$
  and the step-size $\gamma(t)$ satisfy Assumption~\ref{ass:step-size}.
  
  Let $\{ \sx{i}(t), \srho{i}(t) \}$, $i\in \until{N}$, be a sequence generated 
  by the \algname/ distributed algorithm.
  Then, the sequence $\{ \sum_{i=1}^N \big( f_i ( {\sx{i}}(t) ) + M\1^\top \srho{i}(t) \big) \}$ 
  converges to the optimal cost $f^\star$ of~\eqref{eq:primal_problem}.

\label{thm:convergence}  
\end{theorem}
%


\section{Numerical Simulations}
\label{sec:simulations}

In this section we propose a numerical example in which we show the effectiveness 
of the proposed method. 
We consider the following quadratic optimization problem
\begin{align*}
\begin{split}
  \min_{x_1,\ldots,x_N } \: & \: \textstyle\sum\limits_{i=1}^N w_i x_i^2 + r_i x_i
  \\
  \subj \: & \: \ell_i \leq x_i \le u_i, \hspace{1.5cm} i \in\until{N}
  \\
  & \: \textstyle\sum\limits_{i=1}^N (a_i x_i - b_i) \leq 0
\end{split}
\end{align*}
with decision variables $x_i\in\real$, $i\in\until{N}$.  We randomly generated
(positive) $w_i\in\real$ in $[1,20]$ and set $r_i=-20 w_i$.  Moreover, the local
constraint sets are $X_i = [\ell_i, u_i]$, with the extremes uniformly randomly
generated in $[-35,-30]$ and $[30,35]$, respectively. Finally, the coupling
constraint is linear with $a_i$ and $b_i$ randomly generated in $[1,11]$ and
$[0,10]$ respectively.
We set $M=1200$, which turns out to be large enough to contain a dual solution.

In the proposed numerical example we consider a network of $N=20$ agents 
communicating according to an undirected connected Erd\H{o}s-R\'enyi random graph 
$\GG$ with parameter $0.2$. We used a diminishing step-size sequence in the 
form $\gamma(t) = \frac{1}{2}t^{-0.8}$, which satisfies Assumption~\ref{ass:step-size}.

\begin{figure}[!htbp]
\centering
  \includegraphics[scale=0.90]{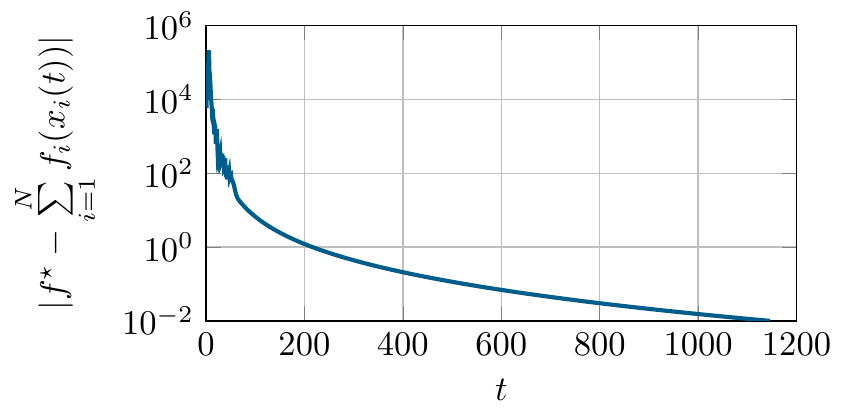}
  %
  \vspace*{-0.4cm}
  \caption{
    Evolution of the cost error, in logarithmic scale.
    }
  \label{fig:cost}
\end{figure}

In Figure~\ref{fig:cost} it is shown the convergence rate of the
distributed algorithm, i.e., the difference between the centralized optimal
cost  $f^\star$ and the sum of the local costs $\sum_{i=1}^N f_i(x_i (t))$,
in logarithmic scale.  It can be seen that the proposed algorithm converges to
the optimal cost with a sublinear rate $O(1/\sqrt{t})$ as expected for a
subgradient method. Notice that the cost error is not monotone since the
subgradient algorithm is not a descent method. 

\begin{figure}[!htbp]
  \centering
  \includegraphics[scale=0.90]{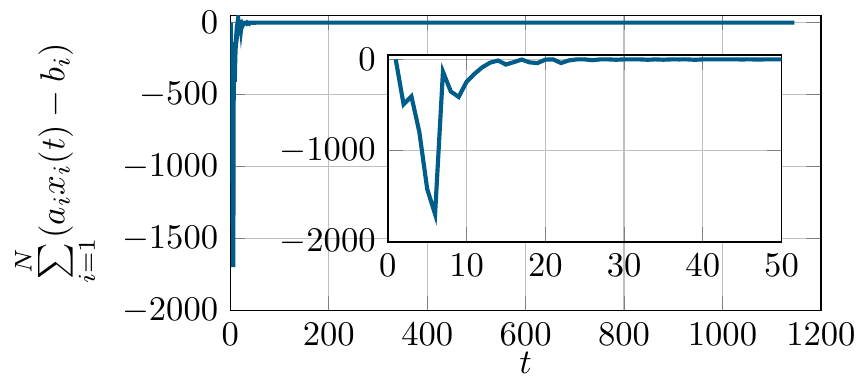}
  %
  \vspace*{-0.4cm}
  \caption{
    Primal violation evolution of $x_i (t)$, $i\in\until{N}$.
  }
  \label{fig:primal_violations}
\end{figure}

In Figure~\ref{fig:primal_violations} we show the violation of the coupling
constraint at each iteration $t$. It is interesting to notice that the violation
asymptotically goes to a nonpositive value. 

\begin{figure}[!htbp]
  \centering
  \includegraphics[scale=0.90]{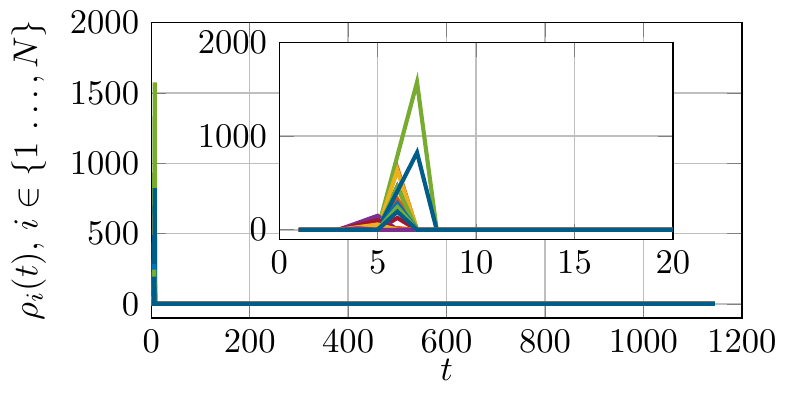}
  %
  \vspace*{-0.4cm}
  \caption{
    Evolution of $\rho_i (t)$, $i\in\until{N}$.
  }
  \label{fig:primal_rho}
\end{figure}

In Figure~\ref{fig:primal_rho} we show the behavior of $\rho_i (t)$,
$i\in\until{N}$, which become zero after some initial iterations, as highlighted
in the zoom.  This behavior reveals that primal local problems at those
iterations would be unfeasible if the relaxation were not present.  Thus, the
``non-relaxed approach'' discussed in Section~\ref{sec:towards} would not work
for this particular numerical example, showing the strength of our relaxed
approach.


\section{Conclusions}
\label{sec:conclusions}
In this paper we have proposed a novel distributed method to solve convex
optimization problems with separable cost function and coupling
constraints. While the algorithm has a very simple structure (a local
minimization and a linear update) its derivation involves a relaxation approach
and a deep tour into duality theory. After a constructive derivation of the
algorithm, we have proven its convergence. Simulations have corroborated the
theoretical results and shown how a first tentative approach, without the
relaxation, would not guarantee convergence a priori.



\begin{small}
  \bibliography{distributed_dual_dual_IFAC}
\end{small}

\end{document}